\documentclass[envcountsame,envcountsect]{llncs}

\newenvironment{proofWithQed}{\par\noindent {\it Proof.} \rm}{\ ~~~\qed}
\newenvironment{proof2}[1]{\par\noindent {\it Proof of #1.} \rm}{ \
~~~\qed } 

\begin{document}
\setcounter{page}{0}

\thispagestyle{empty}
\begin{center}
LaRIA~: Laboratoire de Recherche en Informatique d'Amiens\\
Universit\'e de Picardie Jules Verne -- CNRS FRE 2733\\
33, rue Saint Leu, 80039 Amiens cedex 01, France\\
Tel : (+33)[0]3 22 82 88 77\\
Fax : (+33)[0]03 22 82 54 12\\
\underline{http://www.laria.u-picardie.fr}
\end{center}

\vspace{7cm}

\begin{center}
\parbox[t][5.9cm][t]{10cm}
{\center

{\bf A local balance property of episturmian words

\medskip

G. Richomme$^{\rm a}$
}

\bigskip

\textbf{{L}aRIA \textbf{R}ESEARCH \textbf{R}EPORT~: LRR 2007-02}\\
(Version 2 -- February 2008)
}
\end{center}

\vfill

\hrule depth 1pt \relax

\medskip

\noindent
$^a$ LaRIA, Universit\'e de Picardie Jules Verne, gwenael.richomme@u-picardie.fr\\

\vspace{-2cm}
\pagebreak

\title{A local balance property of episturmian words
}
\author{Gw\'ena\"el Richomme}

\institute{
  UPJV, LaRIA, \\
  33, Rue du Moulin Neuf\\
  80039 Amiens cedex 01\\
  \email{gwenael.richomme@u-picardie.fr}, \\
  \texttt{http://www.laria.u-picardie.fr/{\small $\sim$}richomme/}
}

\date{\today}

\maketitle

\begin{abstract}
We prove that episturmian words and Arnoux-Rauzy sequences can be
characterized using a local balance property. We also give a new
characterization of epistandard words.
\end{abstract}


\medskip

\noindent
\textbf{Important remark:} The first version of this LaRIA Research 
Report 2007-02 contained an error in the proof of a result stating the non-context-freeness of the complement of the set of finite episturmian word. This result, its proof and some comments about it have been removed in this second version of the report. A slightly revised version of the text of this second version was published in ''T.~Harju, J.~Karhum\"aki, and A. Lepist\"o (Eds.), Proceedings of DLT 2007, LNCS 4588, pp 371-381, Springer-Verlag Berlin 
Heidelberg 2007.'' (Thanks to the referees to have seen the above mentioned error)

\section{Introduction}

M. Morse and G.A. Hedlund \cite{MorseHedlund1940} were the firsts to
study in depth a family of words called Sturmian words.  Now a
large litterature exists on these words for which have been proved
numerous characterizations more fascinating the ones than the others
(see for instance \cite{AlloucheShallit2003,BerstelSeebold2002,Pytheas2002}).

Sturmian words are defined over a binary alphabet.  From their various
characteristic properties, some generalizations of Sturmian words have
emerged over larger alphabets. One of them, the so-called Arnoux-Rauzy
sequences, is based on the notion of complexity of a word and is
interesting by its geometrical, arithmetic, ergodic and combinatorial
aspects (see for instance \cite{Pytheas2002}). 

One of the first properties of Sturmian words stated by M. Morse and
G.A. Hedlund \cite{MorseHedlund1940} is the balance property: any
infinite word $w$ over the alphabet $\{a, b\}$ is Sturmian if and only
if it is non-ultimately periodic and balanced, that is the number of
occurrences of the letter $a$ differs in two factors of same length of
$w$ by at most one.  Generalizations of these words were studied for
instance by P.~Hubert \cite{Hubert2000} (see also \cite{Vuillon2003}
for a survey of this property). J.~Justin and L.~Vuillon have stated a
non-characteristic kind of balance property \cite{JustinVuillon2000}
for Arnoux-Rauzy sequences.  Although it was first conjectured that
Arnoux-Rauzy sequences are balanced \cite{DroubayJustinPirillo2001},
J.~Cassaigne, S.~Ferenczi and L.Q.~Zamboni have proved that this does not
necessarily hold \cite{CassaigneFerencziZamboni2000}.

In 1973, E.M.~Coven and G.A.~Hedlund \cite{CovenHedlund1973} stated
that a word $w$ over $\{a, b\}$ is not balanced if and only if there
exists a palindrome $t$ such that $ata$ and $btb$ are both factors of
$w$. This could be seen as a local balance property of Sturmian words
since to check the balance property we do not have to compare all
factors of the same length but only factors on the sets $AtA$ for $t$
factors of $w$.  The previous property can be rephrased in: an
infinite word $w$ over the alphabet $A = \{a, b\}$ is Sturmian if and
only if it is non-ultimately periodic and for any factor $t$ of $w$,
the set of factors belonging to $AtA$ is a subset of $atA \cup Ata$ or
a subset of $btA \cup Atb$. In Section~\ref{secLocalProperty}, we show
that this result can be generalized to Arnoux-Rauzy sequences
contrarily to the balance property.

Actually our result concerns a larger family of infinite words
presented in Section~\ref{secEpisturmian}. Based on ideas of A.~de Luca
\cite{deLuca1997}, Episturmian words were proposed by X.~Droubay,
J.~Justin and G.~Pirillo \cite{DroubayJustinPirillo2001} as a
generalization of Sturmian words. They have observed that Arnoux-Rauzy words
are special episturmian words they called strict episturmian
words. In the binary case episturmian words are the Sturmian words and
the balanced periodic infinite words. Let us note that the case of
remaining balanced words, namely the skew ones, have recently
been generalized \cite{Glen2006,GlenJustinPirillo2006}.

In \cite{DroubayJustinPirillo2001}, episturmian words are defined as
an extension to standard episturmian words (Here we will call {\it
epistandard} these standard episturmian words) previously introduced
as a generalization of standard Sturmian words. In
Section~\ref{secEpistandard}, we generalize to epistandard words a
characterization of standard words proving a converse of a theorem
in \cite{JustinPirillo2002} and stating that an infinite word $w$ is
epistandard if and only if there exists at least two letters such that
$aw$ and $bw$ are both episturmian. The interested reader can also
consult \cite{GlenJustinPirillo2006} and its references for other
characterizations of episturmian words using left extension in the
context of an ordered alphabet.

Our last section comes back to the generalization of the local balance
property introduced by E.M.~Coven and G.A.~Hedlund. One another way to
rephrase it is: an infinite word $w$ over the alphabet $A = \{a, b\}$
is Sturmian if and only if it is non-ultimately periodic and for any
factor $t$ of $w$, the set of factors belonging to $AtA$ is
balanced. This yields a new family of words on which we give partial results.

\section{\label{secEpisturmian}Episturmian and epistandard words}

Even if we assume the reader is familiar with combinatorics on words
(see, e.g., \cite{Lothaire1983}), we precise our notations.  Given an
alphabet A (a finite non-empty set of letters), $A^*$ is the set of
finite words over $A$ including the empty word $\varepsilon$.  The
length of a word $w$ is denoted by $|w|$ and the number of occurrences
of a letter $a$ in $w$ is denoted by $|w|_a$.  The {\it mirror image}
of a finite word $w = w_1 \ldots w_n$ ($w_i \in A$, for $i = 1,
\ldots, n$) is the word $w_n \ldots w_1$ (the mirror image of
$\varepsilon$ is $\varepsilon$ itself). A word equals to its mirror
image is a {\it palindrome}. A word $u$ is a \textit{factor} of $w$ if
there exist words $p$ and $s$ such that $w = pus$. If $p =
\varepsilon$ (resp. $s = \varepsilon$), $u$ is a \textit{prefix}
(resp. \textit{suffix}) of $w$.  A word $u$ is a {\it left special}
(resp. {\it right special}) factor of $w$ if there exist (at least)
two different letters $a$ and $b$ such that $au$ and $bu$ (resp. $ua$
and $ub$) are factors of $w$. A \textit{bispecial factor} is any word
which is both a left and a right special factor (see, e.g.,
\cite{Cassaigne1997}) for more informations on special factors). The
set of factors of a word $w$ will be denoted $Fact(w)$.

The previous notions can be extended in a natural way to any infinite
words. Moreover any \textit{ultimately periodic} infinite word can be
written $uv^\omega$ for two finite words $u, v$ ($v \neq
\varepsilon)$: it is then the infinite word obtained concatenating
infinitely often $v$ to $u$. If $u = \varepsilon$, the word is said
\textit{periodic}.

\medskip

A word $w$ is \textit{episturmian} if and only if its set of factors
is closed by mirror image and $w$ contains at most one left (or
equivalently right) special factor of each length.  A word $w$ is
\textit{epistandard Sturmian} or \textit{epistandard}, if $w$ is
episturmian and all its left special factors are prefixes of $w$.  Let
us note that, in \cite{DroubayJustinPirillo2001}, epistandard words
were introduced by several equivalent ways, and then episturmian words
were defined as words having same set of factors than an epistandard
one.

\medskip

The two theorems below recall a very useful property of episturmian
words which is the possibility to decompose infinitely an episturmian
word using some morphisms. This property already seen for Arnoux-Rauzy
sequences in \cite{ArnouxRauzy1991} is related to the notion of S-adic
dynamical system (see, e.g.  \cite{Pytheas2002} for more details).
This property could be useful to get information on the structure of
episturmian words (see for instance 
\cite{BertheHoltonZamboni2006,LeveRichomme2004,LeveRichomme2006,Richomme2006} for some uses in the
binary cases).

Given an alphabet $A$, a \textit{morphism} $f$ on $A$
is an application from $A^*$ to $A^*$ such that $f(uv) = f(u) f(v)$
for any words $u$, $v$ over $A$.  A morphism on $A$ is entirely
defined by the images of elements of $A$.

\textit{Episturmian morphisms} studied in
\cite{JustinPirillo2002,Richomme2003} are the morphisms defined by
composition of the permutation morphisms and the morphisms $L_a$ and
$R_a$ defined, for $a$ a letter, by
$$L_a \left\{
\begin{tabular}{l}
$a \mapsto a$\\
$b \mapsto ab$, if $b \neq a$,\\
\end{tabular}
\right.
\hspace{1cm}
R_a \left\{
\begin{tabular}{l}
$a \mapsto a$\\
$b \mapsto ba$, if $b \neq a$.\\
\end{tabular}\right.
$$
 
\begin{theorem} {\rm \cite{JustinPirillo2002}}
\label{JP1}
An infinite word $w$ is epistandard if and only if
there exist an infinite sequence of infinite words
$(w^{(n)})_{n \geq 0}$ and an infinite sequence of letters
$(x_n)_{n \geq 1}$ such that $w^{(0)} = w$ and for all
$n \geq 1$, $w^{(n-1)} = L_{x_n}(w^{(n)})$.
\end{theorem}

\begin{theorem} {\rm \cite{JustinPirillo2002}}
\label{JP2}
An infinite word $w$ is episturmian if and only if
there exist an infinite sequence of {\em recurrent} infinite words
$(w^{(n)})_{n \geq 0}$ and an infinite sequence of letters
$(x_n)_{n \geq 1}$ such that $w^{(0)} = w$ and for all
$n \geq 0$, $w^{(n-1)} = L_{x_n}(w^{(n)})$ or $w^{(n-1)} = R_{x_n}(w^{(n)})$.

Moreover, $w$ has the same set of factors than the epistandard word
directed by $(x_n)_{n \geq 1}$.
\end{theorem}

The infinite sequence $(x_n)_{n \geq 1}$ which appears in the two
previous theorem is called the \textit{directive word} of $w$ and is
denoted $\Delta(w)$: Actually in terms of \cite{JustinPirillo2002}, it
is the directive word of the epistandard word having the same set of
factors than $w$. Each episturmian word has a unique directive word.

\medskip

It is worth noting that any episturmian word is \textit{recurrent},
that is, each factor of $w$ occurs infinitely often.  An
infinite word $w$ is recurrent if and only if each factor of $w$
occurs at least twice. Equivalently each factor of $w$ occurs at a
non-prefix position. Thus an infinite word $w$ over an alphabet $A$ is
recurrent if and only if for each of its factors $u$ the set $AuA$ (or
simply $Au$) is not empty.

\medskip

We denote as in \cite{JustinPirillo2002} $Ult(w)$ the set of letters
occurring infinitely often in $\Delta(w)$.  For $B$ a subset of the
alphabet, we introduce a new definition: we call \textit{ultimately
$B$-strict episturmian} any episturmian word $w$ for which
$Ult(\Delta(w)) = B$. Of course this notion is related to the notion
of \textit{$B$-strict episturmian} word (see
\cite[def.~2.3]{JustinPirillo2002}) which is a ultimately
$B$-strict episturmian word whose alphabet (the letters occurring in
$w$) is exactly $B$, and which is nothing else than an Arnoux-Rauzy
sequence over $B$.

As shown in \cite{DroubayJustinPirillo2001}, there is a close
relation between the directive word of an episturmian word and its
special words.  Corollary~\ref{cor1} below will show it again for
ultimately strict episturmian words.

\bigskip

Let $w$ be an episturmian word and $\Delta(w) = (x_n)_{n \geq 1}$ its
directive word.  With notations of Theorem~\ref{JP2}, for $n \geq 1$,
we denote $u_{n,w}$ (or simply $u_n$) the word~:

$$u_{n,w} = L_{x_1}(L_{x_{2}}(\ldots (L_{x_{n-1}}(\varepsilon)x_{n-1})
 \ldots)x_{2})x_1$$

When $n = 1$, $u_{n,w} = \varepsilon$. These words play an important
role in the initial definition of episturmian word by palindromic
closure (see \cite[Sec.~2]{JustinPirillo2002}).  In particular, each
$u_n$ is a palindrom (see for instance
\cite[Lem.~2.5]{JustinPirillo2002}). One can also observe that, if
$Ult(\Delta(w))$ contains at least two letters, then each $u_{n}$ is a
bispecial factor of $w$.  Indeed for $n \geq 1$, $u_{n}$ is a prefix
of the epistandard word $s$ directed by $\Delta(w)$ and so, by
definition of an epistandard word, it is a left special factor of $s$
and so of $w$ by Theorem~\ref{JP2}.  Since the set of factors of $w$
is closed by mirror image and since $u_{n}$ is a palindrom, $u_{n}$ is
a right special factor of $w$.  Conversely let us observe that any
bispecial factor of an episturmian word is a palindrom. Indeed if $u$
is a bispecial factor, then $u$ and its mirror image $\tilde{u}$ are
left special factors of an infinite word containing at most one left
special word of length $|u|$.  It follows the construction of an
epistandard word $w$ by palindromic closure
\cite{DroubayJustinPirillo2001}, that the the words $u_{n,w}$ are the
only palindroms prefixes of $w$. From what precedes, we deduce the
following fact that does not seem to have been already quoted in the
literature:

\begin{remark}
\label{remarkBispecials}
For an episturmian word $w$ with directive word $(x_n)_{n \geq 1}$, a
factor $u$ is bispecial if and only if $u = u_{n,w}$ for an integer $n \geq 1$.
\end{remark}

Another result involving the palindroms $u_n$ is:

\begin{theorem}{\rm \cite[Th.~6]{DroubayJustinPirillo2001}}
\label{thDJPth6}
Let $s$ be an epistandard word over the alphabet $A$ with directive
word $\Delta(s) = (x_n)_{n \geq 1}$. For $n \geq 1$ and $x \in A$,
$u_{n,s}x$ (or equivalently $xu_{n,s}$) is a factor of $s$
if and only if $x$ belongs to $\{ x_i \mid i \geq n \}$.
\end{theorem}

By Theorem~\ref{JP2}, an episturmian word $w$ with a directive word
 $\Delta$ has the same set of factors than the epistandard word with
 directive word $\Delta$. Hence the previous theorem is still valid
 for any episturmian word, and we can deduce:

\begin{corollary}
\label{cor1}
Let $w$ be an episturmian word over an alphabet $A$ and let $B
\subseteq A$ be a set containing at least two different letters.
The word $w$ is a ultimately $B$-strict episturmian word if and only if
for an integer $n_0$, each left special factor 
with $|u| \geq
n_0$ verifies $Au \cap Fact(w) = Bu$.

Moreover for each left special factors with 
$|u| < n_0$, $Bu \subseteq Fact(w)$.
\end{corollary}

The restriction on the cardinality of $B$ ($\geq 2$) will be used in
all the rest of the paper. It is needed to have special factors of
arbitrary length.

\section{\label{secLocalProperty}A new characterization of episturmian words}

Now we give our first main result presented in the introduction as a
kind of local characteristic balance property of episturmian words.

\begin{theorem}
\label{th1}
For a recurrent infinite word $w$, the following assertions are equivalent:
\begin{enumerate}
\item $w$ is episturmian;
\item for each factor
$u$ of $w$, a letter $a$ exists
such that $AuA \cap Fact(w) \subseteq a uA \cup Au a$;
\item for each {\em palindromic} factor
$u$ of $w$, a letter $a$ exists
such that $AuA \cap Fact(w) \subseteq a uA \cup Au a$.
\end{enumerate}
\end{theorem}

In the previous theorem, the letter $a$ and the cardinality of the set
$AuA$ depends on $u$. This is shown for instance by the Fibonacci word
(abaababaabaab\ldots), the epistandard word having $(ab)^\omega$ as
director word, for which $A\varepsilon A \cap Fact(w) = \{aa, ab,
ba\}$, $AaA \cap Fact(w) = \{aab, baa\}$, $AbA \cap Fact(w) =
\{aba\}$, $AaaA \cap Fact(w) = \{baab\}$, \ldots

\medskip

\begin{proof2}{Theorem~\ref{th1}}

\noindent
\textit{Proof of $1 \Rightarrow 2$.}  Assume $w$ is episturmian. Since
the result deals only with factors of $w$, and since by
Theorem~\ref{JP2} an episturmian word have the same set of
factors than an epistandard word, without loss of generality we can
assume that $w$ is epistandard. Let $u$ be a factor of $w$.
Property~2 is immediate if $u$ is not a bispecial factor of $w$. If
$u$ is bispecial in $w$, by Remark~\ref{remarkBispecials}, an integer $n
\geq 1$ exists such that $u = u_{n,w}$. Let $\Delta = (x_i)_{i \geq
1}$ be the directive word of $w$, let $s$ (resp. $t$) be the
epistandard word with $(x_i)_{i \geq n}$ (resp. $(x_i)_{i \geq n+1}$)
as directive word and let $a = x_n$. Letters occurring in $t$ are
exactly the letters of the set $B = \{x_i \mid i \geq n+1\}$.  Since
$s = L_{x_n}(t)$, the factors of length 2 in $s$ are the words $a b$
and $ba$ with $b \in B$. By definition of $\Delta$ and $u_{n,w}$, $w =
L_{x_1}(L_{x_2}( \ldots L_{x_{n-1}}(s)\ldots))$ and $u_{n,w} =
L_{x_1}(L_{x_{2}}(\ldots (L_{x_{n-1}}(\varepsilon)x_{n-1})
\ldots)x_{2})x_1$. Hence by an easy induction on $n$, we deduce $AuA
\cap Fact(w) = a uB \cup Bua \subseteq a uA \cup Aua$.

\medskip

\noindent
\textit{Proof of $2 \Rightarrow 1$.}  Assume that, for any factor $u$
of $w$, a letter $a$ exists such that $AuA \cap Fact(w) \subseteq
a uA \cup Au a$.  In particular, considering the empty word,
we deduce that $AA \cap Fact(w) \subseteq a A \cup Aa$ for a
letter $a$. Hence, for an infinite word $x$, $w = L_a(y)$ if
$w$ starts with $a$ and $w = R_a(y)$ otherwise.

Let us prove that for each factor $v$ of $y$, $AvA \cap Fact(w)
\subseteq b vA \cup Av b$ for a letter $b$. We consider $w
= L_a(y)$ (resp. $w = R_a(y)$).  Let $v$ be a factor of $y$
and let $u = L_a(v)a$ (resp. $u = a R_a(v)$). We
observe that for letters $c, d$, the words $c u d$
is a factor of $w$ if and only if $c v d$ is a factor of
$y$. By hypothesis there exists a letter
$b$  such that $AuA \cap Fact(w) \subseteq b uA \cup
Aub$.  Hence $AvA \cap Fact(w)
\subseteq b vA \cup Avb$.

Letting $x_1 = a$ and iterating infinitely the previous step, we
get an infinite sequence of letters $(x_i)_{i \geq 1}$ and an infinite
sequence of words $(w^{(i)})_{i \geq 0}$ such that $w^{(0)} = w$ and
for all $i \geq 1$, $w^{(i-1)} = L_{x_i}(w^{(i)})$ or $w^{(i-1)} =
R_{x_i}(w^{(i)})$. Due to the fact that $w$ is recurrent, each word
$w^{(i)}$ is also recurrent. By Theorem~\ref{JP2}, the word $w$ is
episturmian.

\medskip

The proof of $1 \Leftrightarrow 3$ is similar to the proof of $1
\Leftrightarrow 2$. Actually, $1 \Rightarrow 3$ is a particular case
of $1 \Rightarrow 2$. When proving $3 \Rightarrow 1$, we need to prove
in the inductive step that $u$ is a palindrome if and only if $v$ is a
palindrome. This is stated by
Lemma 2. 5 in \cite{JustinPirillo2002}~: \textit{a word $u$ is a
palindrome if and only the word $L_a(u)a = aR_a(u)$ is a palindrome}.

\end{proof2}

\medskip

We end this section with few remarks concerning results
that can be proved similarly.

\begin{remark} 
\label{rem1}
Since an infinite word $w$ over an
alphabet $A$ is recurrent if and only if for each factor of $w$ the set $AuA$
is not empty, we have: an infinite word is episturmian if and only if
for each (resp. {\em palindromic}) factor $u$ of $w$, $AuA$ is not
empty and a letter $a$ exists such that $AuA \cap Fact(w) \subseteq a
uA \cup Au a$.
\end{remark}

\begin{remark} 
We have already said that Arnoux-Rauzy sequences over an alphabet $A$
are exactly the (ultimately) A-strict episturmian word. One can ask
for a characterization of these words in a quite similar way than
Theorem~\ref{th1}. Corollary~\ref{cor1} can fulfill this purpose. But
the proof of Theorem~\ref{th1} can also be easily reworked to state~:
{\it an episturmian word $w$ over an alphabet $A$ is a ultimately
$B$-strict episturmian word with $B \subseteq A$ if and only if for
all $n \geq 0$, there exists a {\em (resp. palindromic)} word $u$ of
length at least $n$ and a letter $a$ such that $AuA \cap Fact(w) = auB
\cup Bua$.
}
\end{remark}

\begin{remark} 
Another adaptation of the proof of Theorem~\ref{th1} concerns finite
words: {\it a finite word $w$ is a factor of an infinite episturmian
word if and only if for each factor $u$ of $w$, a letter $a$ exists
such that $AuA \cap Fact(w) \subseteq a u A \cup Au a$}.  We let the
reader verify this result. The main difficulty of the proof is that in
the ``if part'', we do not have necessarily $w = L_a(y)$ or $w =
R_a(y)$. But we have one of the four following cases depending on the
fact that $w$ ends or not with $a$: $w = L_a(y)$, $w = a L_a(y)$, or
$wa = L_a(y)$ or $wa = a L_a(y)$. Except in small cases, we have $|y|
< |w|$ and the technique of the proof of Theorem~\ref{th1} can be
applied.
\end{remark}

\section{\label{secEpistandard}A characterization of epistandard words}

Let us note that for any episturmian word $w$, there exists at least
one letter $a$ such that $a w$ is also episturmian.  Indeed,
since any episturmian word is recurrent, for any prefix $p$ of $w$,
there exists a letter $a_p$ such that $a_p p$ is a factor of
$w$. We work with a finite alphabet hence an infinity of letters
$a_p$ are mutually equal: there exists a letter $a$ such
that $a p$ is a factor of $p$ for an infinity of prefixes (and so
for all prefixes) of $w$. The word $aw$ has the same set of factors
than $w$: it is episturmian.

In restriction to epistandard words, a more precise
result is already know:

\begin{theorem}{\rm \cite[Th.~3.17]{JustinPirillo2002}}
\label{thJP3.17}
If a word $s$ is epistandard, then for each letter $a$ in 
$Ult(\Delta(s))$, $a s$ is episturmian.
\end{theorem}

Up we know the converse of this result has already been stated only in
the Sturmian case (see \cite[Prop. 2.1.22]{BerstelSeebold2002}):
\textit{For every Sturmian word $w$ over $\{a, b\}$, $w$ is standard
episturmian if and only if $aw$ and $bw$ are both Sturmian}.  We
generalize here this result, proving a converse to
Theorem~\ref{thJP3.17}.

\begin{proposition}
\label{gen1}
A {\it non-periodic} word $w$ is epistandard if and only if, for (at
least) two different letters $a$ and $b$,
$a w$ and $b w$ are episturmian.
\end{proposition}

\begin{proofWithQed}
Let $w$ be a non-periodic epistandard word $w$. By 
\cite[Th. 3]{DroubayJustinPirillo2001},
we know that $Ult(\Delta(w))$ contains at least two different letters, say 
$a$ and $b$. By Theorem~\ref{thJP3.17}, 
$a w$ and $b w$ are episturmian.

\medskip

Assume now that for two different letters $a$ and $b$, $a w$ and $b w$
are episturmian. Since $a w$ (and also $b w$) is recurrent, $w$ has the
same set of factors than $a w$ and so $w$ is episturmian.  Moreover
each prefix $p$ is left special (since $a p$ and $b p$ are factors of
$w$). Since any episturmian word has at most one left special factor for each
length, the left special factors of $w$ are its prefixes: $w$ is
epistandard.
\end{proofWithQed}

\medskip

Let us give a more precise result:

\begin{theorem}
\label{mainTh}
Let $w$ be an infinite word over the alphabet $A$ and assume $B
\subseteq A$ contains at least two different letters. The two
following assertions are equivalent:
\begin{enumerate}
\item The word $w$ is ultimately $B$-strict epistandard;
\item For each letter $a$ in $A$, 
$a w$ is episturmian if and only if $a$ belongs to $B$.
\end{enumerate}
\end{theorem}

\begin{proofWithQed}
Assume first that $w$ is $B$-strict epistandard, that is,
$Ult(\Delta(w)) = B$.  By Theorem~\ref{thJP3.17}, for each letter
$a$ in $B$, $a w$ is episturmian. For any integer $n \geq
0$, the word $u_{n, w}$ is a prefix of $w$.  If $a$
does not belong to $B$, by Theorem~\ref{thDJPth6}, for at least one
integer $n \geq 0$, $au_{n,w}$ is not a factor of $w$.  Thus the word
$a w$ is not recurrent and so it is not episturmian. Hence if $w$ is
$B$-strict epistandard, for each letter $a$ in $A$, $a w$ is
episturmian if and only if $a$ belongs to $B$.

\medskip

Assume now that for each letter $a$ in $A$, $a w$ is episturmian if
and only if $a$ belongs to $B$. Since $B$ contains at least two
letters, by Proposition~\ref{gen1}, $w$ is epistandard.  As a
consequence of Theorem~\ref{thDJPth6}, we can deduce $Ult(\Delta(w)) =
B$.
\end{proofWithQed}

\section{A new family of words}

In this section, we consider recurrent infinite words $w$ over an
alphabet $A$ having the following property:

\newcommand{\Pp}{${\cal P}$ }
\newcommand{\Ppf}{${\cal P}$}
\begin{description}
\item{Property \Ppf:} for any word $u$ over $A$, the set of factors of
$w$ belonging to $AuA$ is balanced,\\
that is, for any word $u$ and for any letters $a, b, c,
d$, if $aub$ and $cud$ are factors of $w$ then $\{a, b\} \cap \{c, d\}
\neq \emptyset$.
\end{description}

Any word verifying Assertion 2 in Theorem~\ref{th1} also verifies
Property \Ppf. As shown by the word $(abc)^\omega$, the converse does not
hold. In other words, any episturmian word verifies Property~\Ppf, but
this is not a characteristic property (except in the binary case for
which it is immediate that a word $w$ verifies Property \Pp if and
only if for all words $u$, $aua$ or $bub$ is not a factor of $w$).

We prove:

\begin{proposition}
\label{prop6.1}
A recurrent word $w$ over an alphabet $A$ verifies property~\Pp if and only if one of the two following assertion holds:
\begin{enumerate}
\item $w$ is episturmian;
\item there exist an episturmian morphism $f$, three different letters
$a, b, c$ in $A$ and a word $w'$ over $\{a,b,c\}$ such that $w =
f(w')$, $w'$ verifies Property~\Pp and the three words $ab$, $bc$ and
$ca$ are factors of $w'$.
\end{enumerate}
\end{proposition}

This proposition is a consequence of the next two lemmas.

\begin{lemma}
\label{lemma6.3}
If a recurrent infinite word $w$ verifies property~\Ppf, then one of
the two following assertion holds:
\begin{enumerate}
\item $w = L_\alpha(w')$ or $w = R_\alpha(w')$ for a letter $\alpha$
and a recurrent infinite word $w'$;

\item there exist three different letters $a, b, c$ such that $w \in
\{a, b, c\}^\omega$ and the three words $ab$, $bc$ and $ca$ are
factors of $w$.
\end{enumerate}
\end{lemma}

\begin{proofWithQed}
We first observe that if $AA \cap Fact(w) \subseteq \alpha A \cup A
\alpha$ then (as in the proof of Theorem~\ref{mainTh}) $w =
L_\alpha(w')$ or $w = R_\alpha(w')$, for a letter $\alpha$ and a
recurrent infinite word $w'$.

We assume from now on that $AA \cap Fact(w) \not\subseteq \alpha A
\cup A \alpha$. 

For any letter $\alpha$ in $A$, $\alpha \alpha$ is not a factor of
$w$.  Indeed if such a word is a factor of $w$, then, for any factor
$\beta \gamma$ with $\beta$ and $\gamma$ letters, by Property~\Ppf, $\beta
= \alpha$ or $\gamma = \alpha$, that is $AA \cap Fact(w) \subseteq
\alpha A \cup A \alpha$.

The alphabet $A$ contains at least three letters. Indeed if $A$
contains at most two letters $a$ and $b$, then Property~\Pp implies
that $aa$ and $bb$ are not simultaneously factors of $w$, and so we
have $AA \cap Fact(w) \subseteq aA \cup Aa$ or $AA \cap Fact(w)
\subseteq bA \cup Ab$.

Let us prove that $A$ contains exactly three letters.  Assume by
contradiction that $A$ contains at least four letters. Let $a$
(resp. $b$) be the first (resp. the second) letter of $w$. Since $aa$
is not a factor of $w$, $a \neq b$. At least two other letters $c$ and
$d$ occur in $w$ ($c, d \not\in \{a, b\}$, $c \neq d$).  By
Property~\Ppf, each occurrence of $c$ is preceded by $a$ or by $b$.
Assume that $ac$ occurs in $w$.  Since $ab$ also occurs, for any
letter $\alpha$ not in $\{a, b, c\}$, each occurrence of $\alpha$ is
preceded and followed by the letter $a$. But $AA \cap Fact(w) \not\subseteq a A
\cup A a$. Hence $bc$ or $cb$ occurs in $w$. But then the factor $ad$
contradicts Property~\Ppf.
Assume now that $bc$ occurs in $w$. Since $ab$ also occurs, for any
letter $\alpha$ not in $\{a, b, c\}$, each occurrence of $\alpha$ is
preceded and followed by $b$. But $AA \cap Fact(w) \not\subseteq b A
\cup A b$. Hence $ac$ or $ca$ occurs in $w$. But then the factor $db$
contradicts Property~\Ppf.

Until now we have proved that $w$ is written on a three-letter
alphabet and contains no word $\alpha \alpha$ with $\alpha$ a letter.
Assume that, for two letters $a$ and $b$, $ab$ is a factor of $w$ but
not $ba$. Then for an integer $n \geq 1$, $a(bc)^na$ (let recall that
$aa$, $bb$, $cc$ and $ba$ are not factors of $w$), and so $ab$, $bc$
and $ca$ are factors of $w$. Now if, for all letters $\alpha$ and
$\beta$, $\alpha\beta$ and $\beta\alpha$ are factors of $w$ then
denoting $a$, $b$ and $c$ the letters occurring in $w$, once again
$ab$, $bc$ and $ca$ are factors of $w$.
\end{proofWithQed}

\medskip

\begin{lemma}
\label{lemma6.2}
Let $\alpha$ be a letter, $w$ and $w'$ be recurrent words such that $w
= L_\alpha(w')$ or $w = R_\alpha(w')$. The word $w$ verifies
Property~\Pp if and only if $w'$ verifies Property~\Ppf.
\end{lemma}

\begin{proofWithQed}
We first assume $w = L_\alpha(w')$.

Assume that $w$ does not verify Property~\Ppf: $aub$ and $cud$ are
factors of $w$ for some letters $a, b, c, d$ and a word $u$ such that
$\{a, b\} \cap \{c, d\} = \emptyset$. At least one of the two letters
$a$ and $b$ is different from $\alpha$ and at least one of the two
letters $c$ and $d$ is different from $\alpha$. Since $w =
L_\alpha(w')$, we deduce that $u \neq \varepsilon$, and that $u$
begins and ends with $\alpha$: $u = L_\alpha(v)\alpha$ for a word
$v$. Thus $aub = aL_\alpha(v)\alpha b$ and $cud = cL_\alpha(v)\alpha
d$. We observe that if $a \neq \alpha$ (resp. $c \neq \alpha$),
$\alpha aL_\alpha(v)\alpha b$ (resp.  $\alpha cL_\alpha(v)\alpha d$)
is a factor of $w$. Thus we can deduce that $avb$ and $cvd$ are
factors of $w'$ (even if one of the letters $a, b, c, d$ is $\alpha$):
the word $w'$ does not verify Property~\Ppf.

Assume conversely that the word $w'$ does not verify Property~\Ppf:
$aub$ and $cud$ are factors of $w'$ for some letters $a, b, c, d$ and
a word $u$ such that $\{a, b\} \cap \{c, d\} = \emptyset$.  The word
$aL_\alpha(u)\alpha b$ is a factor of $w$ (if $b = \alpha$, this is
still true since we work with infinite words and so in this case
$au\alpha b'$ is a factor of $w$ for a letter $b'$). Similarly $cL_\alpha(u)\alpha d$ is
a factor of $w$: the word $w$ does not verify
Property~\Ppf.

\medskip

The proof when $w = R_\alpha(w')$ is similar. Note that the fact that
$w'$ is recurrent is needed for the last part of the proof to know
when $a = \alpha$, that $a' \alpha u b$ is a factor of $w'$ for a
letter $a'$.
\end{proofWithQed}

\medskip

\begin{proof2}{Proposition~\ref{prop6.1}}
Assume $w$ is a recurrent word that verifies Property~\Pp but that does not
verifies Assertion~2 of Lemma~\ref{lemma6.3}. Then $w = L_\alpha(w')$
or $w = R_\alpha(w')$, with $w'$ a recurrent word. By
Lemma~\ref{lemma6.2}, $w'$ verifies Property~\Ppf. 

Thus using Lemmas~\ref{lemma6.3} and \ref{lemma6.2}, we can prove by induction
that, for any integer $n \geq 0$, one of the two following assertions holds~:
\begin{itemize}
\item there exist recurrent infinite word $w^{(0)} = w$, $w^{(1)}$,
\ldots $w^{(n)}$, and letters $a_1$, \ldots, $a_n$ such that for each
$1 \leq p \leq n$, $w^{(p-1)} = L_{a_p}(w^{(p)})$ or $w^{(p-1)} =
R_{a_p}(w^{(p)})$, and $w^{n}$ verifies property~\Ppf;
\item for an integer $m \leq n$, 
there exist recurrent infinite word $w^{(0)} = w$, $w^{(1)}$, \ldots
$w^{(m)}$, and letters $a_1$, \ldots, $a_m$ such that for each $1 \leq p
\leq m$, $w^{(p-1)} = L_{a_p}(w^{(p)})$ or $w^{(p-1)} = R_{a_p}(w^{(p)})$, 
and $w^{(m)}$ verifies Assertion~2 of Lemma~\ref{lemma6.3}.
\end{itemize}
Hence the proposition is a consequence of Theorem~\ref{JP2}.
\end{proof2}

\section{Conclusion}

The reader has certainly noticed that words verifying Property~\Pp are
not completely characterized. For this, one should have to better know
ternary recurrent words verifying Property~\Pp and containing the
words $ab$, $bc$ and $ca$ as factors.

Let us give examples of such words. One can immediately verify that if
$ab$, $bc$ and $ca$ are the only words of length 2 that are factors of
a word $w$, then $w$ is $(abc)^\omega$, $(bca)^\omega$ or
$(cab)^\omega$. When a recurrent word $w$ verifying property~\Pp has
exactly the words $ab$, $bc$, $ca$ and $ba$ as factors of length 2,
one can see that $w$ is a suffix of a word $f(w')$ where $w'$ is a
Sturmian word over $\{a, b\}$ and $f$ is the morphism defined by $f(a)
= (ab)^nc$ and $f(b) = (ab)^{n+1}c$ for an integer $n \geq 1$.  When
$f$ is replaced by one of the following morphisms $g_1$ or $g_2$, we
can get other examples of ternary words verifying Property~\Pp (and
containing exactly 5 factors of length 2 with amongst them $ab$, $bc$
and $ca$)~: $g_1(a) = (ab)^nc$, $g_1(b) = (ab)^{n}cb$, $g_2(a) =
(ab)^nc$, $g_2(b) = (ab)^{n+1}cb$.  Our final example is the periodic
word $(abcabacbabcb)^\omega$ which verifies Property~\Pp and contains
as factors all words of length 2 except $aa$, $bb$, $cc$: this word
could be seen as the morphic image of $a^\omega$ by the morphism that
maps $a$ onto $abcabacbabcb$.

All these examples lead to the question: Are all ternary recurrent
words verifying Property~\Pp and containing $ab$, $bc$ and $ca$ as
factors are suffix of a word $f(w')$ with $w'$ a recurrent balanced
word (that is a Sturmian word or a periodic balanced word) and with
$f$ a morphism?  If it is true, which are the possible values for $f$?

\medskip

\noindent
\textbf{Acknowledgements.}  The author would like to thanks
J.-P. Allouche for his questions that have initiated the present work.

\end{document}